\documentclass[]{revtex4}
\usepackage{graphicx}

\begin{document}

\draft
\title{Quantum three body problems using harmonic oscillator bases
with different sizes}
\author{B. Silvestre-Brac%
, R. Bonnaz%
 \\
Institut des Sciences Nucl\'{e}aires, IN2P3, CNRS, Universit\'{e}
Joseph Fourier, \\ Av. des Martyrs 53, F-38026 Grenoble-Cedex, France
\\C. Semay%
, F. Brau%
\\
Universit\'{e} de Mons-Hainaut, Place du Parc 20,
B-7000 Mons, Belgium}

\begin{abstract}
{\small We propose a new treatment for the quantum three-body problem.
It is
based on an expansion of the wave function on harmonic oscillator
functions with different sizes in the Jacobi coordinates. The matrix
elements of the Hamiltonian can be calculated without any
approximation and the precision is restricted only by the dimension of
the basis. This method can be applied whatever the system under
consideration. In some cases, the convergence property is greatly
improved in this new scheme as compared to the old traditional method.
Some numerical tricks to reduce computer time are also presented.}
\end{abstract}

\pacs{12.39.Pn, 14.20.-c}


\maketitle

\section{Introduction}
\label{sec:intro}

The quantum problem of three interacting particles is a very old one,
since it is present in very different domains of physics: molecular,
atomic, nuclear and hadronic physics among others. Besides the fact
that there exists a large number of such systems in nature, it is
interesting because it is much more difficult to solve than the
relatively easy two-body problem. One difficulty comes from the fact
that statistical approaches or even many-body technics are not
efficient for three-body problems; in particular a good treatment of
the center of mass is necessary and internal coordinates must be
employed. Another difficulty appears if the three particles are
identical; in that case one must fulfill the Pauli principle which is
not easy to manage with internal coordinates.

There exists a lot of different technics to solve the three-body
problem; let us cite, among others, quantum Monte Carlo method
\cite{carl,lind,ham}, Faddeev equations \cite{gloc}, hyperspherical
formalism \cite{flr}, stochastic variational method \cite{suz},
expansion on orthogonal bases, for example harmonic oscillator (OH)
\cite{isg,sil96}. In principle, all these methods tend to the exact
result
if some parameters (number of states, number of amplitudes, number of
mesh points,...) tend to infinity. The convergence properties depend
not only on the type of method, but also on the system and the
dynamics themselves. Of course, one searches the minimum of
computational effort for a given precision. Each method has its own
advantages and its own drawbacks. For example, dealing with a
semirelativistic kinematics is not an easy task with Faddeev or
hyperspherical formalism on a mesh, hard core or very short range
repulsive
potentials are very difficult to implement with the stochastic
variational method or the HO basis.

The aim of this paper is to revisit the HO method to accelerate the
convergence of the results. So we want to obtain the same precision
with an expansion needing less quanta, and thus less basis states.
Besides the fact that a smaller number of quanta means less storing
memory and less computational time, it has also the advantage to give
a more physical idea of the wave function. Indeed if we obtain a good
wave function with, let say, $N=1000$ basis states the physical
interpretation of this wave function is difficult; on the other hand
if we get the same precision with $N=10$ basis states, one grasps
better the physical contents of the system because the degrees of
freedom chosen are more adapted to it. Moreover, if we are interested
by some
observable built from this wave function the gain is even more
impressive. To obtain the average value of the operator on the
calculated wave function, the first case needs to compute one million
terms, whereas one hundred terms are enough in the second case.

The traditional approach based on HO basis considers that the harmonic
oscillator wave function have the same size (or the same scale) in
both Jacobi coordinates. This was an unavoidable requirement to
calculate rapidly and precisely the matrix elements of the potential
within this basis. If the three particles have the same mass, as in
most problems of
nuclear physics, this condition is not a flaw; but if the particles
have very different masses this condition is not well suited because
the basis states can hardly reproduce the physical asymmetry. This
paper presents a method to deal with this asymmetry by using HO wave
functions with different sizes for different Jacobi coordinates.
One can be very skeptical on the possibility to calculate
rigorously the matrix elements in such a basis because it is known
that
expanding a HO wave function with one size on a basis of HO wave
functions with another size requires an infinite number of terms.
However, we will show that we perfectly achieve this goal if we
define correct changes of variables. This possibility opens the door
to a better convergence of the method.

We are aware that the present method cannot compete with more
sophisticated technics to obtain a very precise result close to the
exact one; but the price to pay is also much less. So, we believe that
it is a good compromise between computational and technical
difficulties and precision of the results. A serious advantage of this
approach is that the use of a relativistic kinematics is not a
problem; the Fourier transform of a HO function is again a HO function
so that the matrix elements of the operator are easily calculated in
momentum representation. Another interesting advantage of using a HO
basis is its universality; it is systematic so that dealing with
orbitally or radially excited states is of the same
difficulty than dealing with the ground state. This is not the case
for most of other methods.

The paper is organized as follows. In the next section, we develop the
theory with special accent put on the differences with the traditional
formalism. In Sec.~\ref{sec:num} we present some numerical aspects
that
allowed us to gain comfortable computer time. In Sec.~\ref{sec:res} a
detailed analysis of convergence properties, as well as a simple
application are discussed. In the last section the
conclusions are drawn. Some very technical details are
relegated in the appendices.

\section{Theory}
\subsection{Jacobi coordinates}
\label{sec:theory}

Each particle $i$ ($i$=1, 2, 3) is characterized by a mass $m_i$ and
by
various dynamical degrees of freedom: internal degrees of freedom
symbolized generically by $\alpha_i$ and by its position ${\bf r}_i$
in a given reference frame. In case of electrons for atomic physics
$\alpha_i$ stands for spin and the corresponding magnetic number; in
the
case of nucleons for nuclear physics $\alpha_i$ includes, in addition
to spin, isospin degrees of freedom, while in the case of quarks for
subnuclear physics $\alpha_i$ includes, in addition to spin and
isospin, color degrees of freedom. The conjugate momentum of
${\bf r}_i$ is denoted ${\bf p}_i$. Let us define by $m$ an arbitrary
reference mass and the dimensionless parameters $\omega_i=m_i/m$,
 $\omega_{ij}=\omega_i+\omega_j$, $\omega=\omega_1+\omega_2+\omega_3$.

In order to treat correctly the center of mass motion, it is necessary
to introduce the center of mass position ${\bf R}$ and the total
momentum ${\bf P}$, defined as usual
\begin{equation}
{\bf R}=\frac{\omega_1 {\bf r_1}+\omega_2 {\bf r_2}+\omega_3 {\bf
r_3}}{\omega} \quad ; \quad {\bf P}={\bf p}_1+{\bf p}_2+{\bf p}_3.
\end{equation}
The dimensionless Jacobi coordinates ${\bf x}$ and ${\bf y}$
corresponding to internal relative positions can be defined
with several prescriptions.

For people working with traditional HO functions, the usual
definition is
\begin{equation}
\label{trdjc}
b \, {\bf x}=\sqrt{\frac{2 \omega_2 \omega_3}{\omega_{23}}} \:
({\bf r}_2 -{\bf r}_3) \quad ; \quad b \, {\bf y} =\sqrt{\frac{2
\omega_1
\omega_{23}}{\omega}} \: \left(\frac{\omega_2 {\bf r}_2+\omega_3
{\bf r}_3}{\omega_{23}}-{\bf r}_1\right).
\end{equation}
Here, the scale parameter $b$ implies a unique size for HO functions.
The deep reason for choosing such a precise definition is the
following; when dealing with the potential, one needs to express the
Jacobi coordinates that are derived from a permutation of the
particles. With the choice (\ref{trdjc}) all these Jacobi coordinates
are related by {\em orthogonal transformations}; this nice property
allows to simplify a lot the numerical calculations. The parameter
$b$ can be determined from a variational procedure.

In our approach we introduce two scale parameters, one for each Jacobi
coordinate, so that we define more simply
\begin{equation}
\label{jc}
b_x \, {\bf x}=({\bf r}_2 -{\bf r}_3) \quad ; \quad b_y \, {\bf y} =
\frac{\omega_2 {\bf r}_2+\omega_3 {\bf r}_3}{\omega_{23}}-{\bf r}_1.
\end{equation}
The two parameters $b_x$ and $b_y$ can also be determined by a
variational procedure. For arbitrary values of $b_x$ and $b_y$, both
definitions (\ref{trdjc}) and (\ref{jc}) of the Jacobi coordinates
obviously differ; they are nevertheless identical if we impose the
relationship
\begin{equation}
\label{lien}
b_x=b \sqrt{\frac{\omega_{23}}{2 \omega_2 \omega_3}} \quad {\rm and}
\quad
b_y=b \sqrt{\frac{\omega}{2 \omega_1 \omega_{23}}}.
\end{equation}
Thus, our new theory must coincide with the old one if we maintain the
conditions (\ref{lien}). This is a drastic check for our numerical
codes.

The conjugate momenta corresponding to ${\bf x}$ and ${\bf y}$ are
denoted ${\bf p}$ and ${\bf q}$ respectively. Their expression in
terms of ${\bf p}_i$ are straightforward.

\subsection{Basis states}

With those definitions (\ref{jc}), particle 1 plays a special role and
the
natural coupling is [1(23)]. Nevertheless we have the freedom to
choose the particle order. If we were able to perform a rigorous
treatment (number of quanta infinite), this order would be irrelevant;
however the expansion is truncated, the order makes a difference,
and there exists a special order which gives better results. We will
show an example later.

The total wave function $\Psi$ is expanded on basis states
\begin{equation}
\label{fot}
\Psi({\bf x},{\bf y})=\sum_{i=1}^{N} \eta_{\sigma_i}
\Phi_{k_i}({\bf x},{\bf y}).
\end{equation}
In this expression $\eta_{\sigma_i}$ is the part of the wave function
corresponding to the internal degrees of freedom (spin, isospin,
color)
and means symbolically $[\alpha_1(\alpha_2
\alpha_3)_\sigma]_{\alpha}$; the various
indices $\sigma$ stand for the intermediate couplings and correspond
to a finite number of states. The space part
$\Phi_{k_i}({\bf x},{\bf y})$
is a coupled product of two HO functions
\begin{equation}
\label{foe}
\Phi_k({\bf x},{\bf y})=[\phi_{nl}({\bf x})\phi_{\nu \lambda}({\bf
y})]_L,
\end{equation}
where $n$ ($\nu$) and $l$ ($\lambda$) are the radial and orbital
quantum numbers for the Jacobi
coordinate ${\bf x}$ (${\bf y}$); $L$ is the orbital angular momentum
of the system, and the
index $k$ gathers the quantum numbers ${n,l,\nu,\lambda,L}$. The
functions
$\phi_{nlm}({\bf x})$ = $\frac{u_{nl}(x)}{x}Y_{lm}(\hat{x})$ are the
usual HO wave functions, defined in any textbook on quantum mechanics.
In the following, we will use extensively the matrix elements
\begin{equation}
\label{potoh}
V_{nl,n'l'}(a)=\langle \phi_{nl}({\bf x}) \mid V(ax) \mid
\phi_{n'l'}({\bf x}) \rangle = \delta_{ll'}\int_0^{\infty} u_{nl}(x)
u_{n'l}(x)
V(ax) dx.
\end{equation}
An efficient method to calculate them is discussed later on.

An interesting property of the space functions (\ref{foe}) is the
orthogonality condition
\begin{equation}
\langle \Phi_k({\bf x},{\bf y}) \mid \Phi_{k'}({\bf x},{\bf y})
\rangle = \delta_{kk'},
\end{equation}
which is valid whatever the size parameters.

The number of quanta of the function (\ref{foe}) is simply
$2n+l+2\nu + \lambda$. In the expansion of the total wave function
(\ref{fot}), we always consider {\em all the basis states} (\ref{foe})
with a number of quanta less or equal to a given number $N_Q$. This
prescription is absolutely fundamental to treat correctly the Pauli
principle (see Sec.~\ref{subsec:iden}).

\subsection{The Hamiltonian}
Since our approach is essentially of type ``potential model" the
Hamiltonian takes the traditional form
\begin{equation}
H=T+V.
\end{equation}
We are able to treat equally well both types of kinetic energy
operator $T$, nonrelativistic or semirelativistic.

The nonrelativistic operator is given by
\begin{equation}
\label{tnr}
T=T_{\rm nr}=\sum_{i=1}^3 \frac{{\bf p}_i^2}{2 m_i} - \frac{{\bf P}^2}
{2M} = \frac{{\bf p}^2}{2 \mu_p}+\frac{{\bf q}^2}{2 \mu_q},
\end{equation}
where $M=\omega m$ is the total mass and where $\mu_p$ and $\mu_q$ are
quantities proportional to the reduced masses. They are defined by
\begin{equation}
\mu_p = \frac{mb_x^2 \omega_2 \omega_3}{\omega_{23}} \quad ; \quad
\mu_q = \frac{mb_y^2 \omega_1 \omega_{23}}{\omega}.
\end{equation}

The semirelativistic operator needs to be evaluated in the rest frame
$\bf P=0$; hence, we have
\begin{equation}
T=T_{\rm sr}= \sum_{i=1}^3 \sqrt{{\bf p}_i^2+m_i^2}-M = K_1+K_2+K_3-M.
\end{equation}
In the rest frame the expression of each term is written
\begin{eqnarray}
\label{tsr}
K_1&=& \sqrt{\frac{{\bf q}^2}{b_y^2}+ \omega_1^2 m^2}, \nonumber \\
K_2&=& \sqrt{\left(\frac{\omega_2 {\bf q}}{b_y \, \omega_{23}}+
\frac{{\bf p}}{b_x} \right)^2+\omega_2^2 m^2}, \nonumber \\
K_3&=& \sqrt{\left(\frac{\omega_3 {\bf q}}{b_y \, \omega_{23}}-
\frac{{\bf p}}{b_x} \right)^2+\omega_3^2 m^2}.
\end{eqnarray}

Although it is possible to deal with three-body forces in this
formalism, we consider in this paper only two-body forces
\begin{equation}
V=V_{12}+V_{13}+V_{23},
\end{equation}
where $V_{ij}$ represents the interaction between particle $i$ and
particle $j$. It can be decomposed generally as
\begin{equation}
\label{pot}
V_{ij}= \sum_s \hat{O}_{ij}^{(s)} V_{ij}^{(s)}(\mid {\bf r}_i -
{\bf r}_j \mid),
\end{equation}
where $\hat{O}_{ij}^{(s)}$ is the operator acting in the space of
internal degrees of freedom (spin, isospin, color). There exist in
general several different structures $(s)$ compatible with invariance
symmetries. The most general potential must take care of all these
possibilities by a summation over the various structures. The space
part for a given structure is the form factor
$V_{ij}^{(s)}(\mid {\bf r}_i - {\bf r}_j \mid)$.

\subsection{Matrix elements}
\subsubsection{Brody-Moshinsky coefficients}

The calculation of the matrix elements in the basis (\ref{foe}) is one
of the novelties developed in this paper. It is based on the use of
generalized Brody-Moshinsky (or Smirnov) coefficients (BMC). This
technique was employed long time ago by nuclear physicists but seems
to be not often used nowadays. One can find the interesting
properties of BMC in several textbooks (see for example
\cite{law,bro}) or papers \cite{sil85}. The only thing that is needed
here is that they relate HO functions with arguments that are
transformed by a rotation. More explicitly, we have
\begin{eqnarray}
\label{bmc}
[\phi_{n_1l_1}({\bf r}\cos \beta+{\bf R} \sin \beta) \phi_{n_2l_2}
(- {\bf r}\sin \beta+{\bf R} \cos \beta)]_{\lambda} &=& \nonumber \\
\sum_{n,l,N,L} \langle nlNL;\lambda \mid n_1l_1n_2l_2;\lambda \rangle_
{\beta}[\phi_{nl}({\bf r})\phi_{NL}({\bf R})]_{\lambda}. & &
\end{eqnarray}
The various quantum numbers appearing in the BMC
$\langle nlNL;\lambda \mid n_1l_1n_2l_2;\lambda \rangle_{\beta}$ are
constrained by triangular inequalities, by parity conservation and by
conservation of the number of quanta.

\subsubsection{kinetic energy}
The matrix elements for the non-relativistic operator (\ref{tnr}) are
very well known
\begin{equation}
\label{tnrme}
\langle \Phi_{k_i} \mid T_{\rm nr} \mid \Phi_{k_j} \rangle =
\delta({\bar k}_i,{\bar k}_j) K_p(n_i,n_j) +
\delta({\bar k}_i,{\bar k}_j) K_q(\nu_i,\nu_j),
\end{equation}
where the notation $\delta({\bar k}_i,{\bar k}_j)$ means the Kronecker
symbol for every quantum number of the basis except those appearing in
the matrix element in front of it ($K_{p}$ for the first, $K_{q}$ for
the second).
This last term is given by
\begin{eqnarray}
\label{kp}
K_p(n_i,n_j)&=&\frac{1}{2 \mu_p} \left[(2n_i+l_i+3/2) \delta_{n_i,n_j}
+ \sqrt{n_i(n_i+l_i+1/2)} \delta_{n_i,n_j+1}\right. \nonumber \\
&+&\left. \sqrt{n_j(n_j+l_j+1/2)} \delta_{n_j,n_i+1}\right],
\end{eqnarray}
and an analogous expression for $K_q$.

The matrix elements for the semirelativistic operators (\ref{tsr})
seem more complicated. However, it is convenient to work in momentum
representation. Indeed the Fourier transform of the space function
(\ref{foe}) is exactly of the same form (with an extra phase factor)
with ${\bf p}$ and ${\bf q}$ replacing ${\bf x}$ and ${\bf y}$. The
first term $K_1$ of the operator is thus very easy to calculate
\begin{equation}
\label{ksr1}
\langle \Phi_{k_i} \mid K_{1} \mid \Phi_{k_j} \rangle =
\delta({\bar k}_i,{\bar k}_j) \frac{(-1)^{\nu_i+\nu_j}}{b_y}
{\cal K}_{\nu_i \lambda_i,\nu_j \lambda_j}(b_y m \omega_1),
\end{equation}
where the dynamical ingredient $\cal K$ is reduced to a single
integral
\begin{equation}
\label{calk}
{\cal K}_{nl,NL}(\mu) = \int_0^{\infty} dq \, u_{nl}(q) u_{NL}(q)
\sqrt{q^2+\mu^2}.
\end{equation}

The idea for calculating the matrix elements of $K_2$ relies on a
trick that can be applied with adaptation to the other matrix
elements. We remark that, in $K_2$, the vector present under the
square root, namely $\frac{\omega_2 {\bf q}}{b_y \omega_{23}}+
\frac{\bf p}{b_x}$, can be made proportional to some vector ${\bf r}$
which is obtained from ${\bf p}$ and ${\bf q}$ by a rotation with some
angle $\beta_1$. One introduces the vector ${\bf s}$ orthogonal to
${\bf r}$ and moves, with help of BMC, from (${\bf p}$,${\bf q}$)
representation for HO to the (${\bf r}$,${\bf s}$) representation. The
calculation of the matrix element in this representation is then quite
easy. The final result is
\begin{eqnarray}
\label{ksr2}
\langle \Phi_{k_i} \mid K_{2} \mid \Phi_{k_j} \rangle &=&
\delta({\bar k}_i,{\bar k}_j) \frac{\pi_2}{\eta_1} \sum_
{n_2,l_2,\nu_2,\nu_2',\lambda_2} (-)^{\nu_2+\nu_2'}
\langle \nu_2 \lambda_2 n_2 l_2; L_i
\mid n_j l_j \nu_j \lambda_j;L_i \rangle_{\beta_1} \nonumber \\
&& \langle \nu_2' \lambda_2 n_2 l_2; L_i \mid n_i l_i \nu_i \lambda_i;
L_i \rangle_{\beta_1}{\cal K}_{\nu_2' \lambda_2,\nu_2 \lambda_2}
(\eta_1 m \omega_2).
\end{eqnarray}
The phase is $\pi_2=(-1)^{l_i+l_j}$ and we have introduced geometrical
factors
\begin{eqnarray}
\label{para1}
\alpha_1&=&\frac{\sqrt{b_x^2 \omega_2^2+b_y^2 \omega_{23}^2}}
{\omega_{23}},  \nonumber \\
\eta_1&=&\frac{b_x b_y}{\alpha_1},
\end{eqnarray}
and the angle for the rotation
\begin{equation}
\label{beta1}
\cos \beta_1 = \frac{b_y}{\alpha_1} \quad ; \quad \sin \beta_1 = \frac
{b_x \, \omega_2}{\alpha_1 \, \omega_{23}}.
\end{equation}

The matrix element for the $K_3$ term is obtained exactly with the
same trick. The result can be obtained from (\ref{ksr2}) with the
phase $\pi_3=1$, with other geometrical factors $\alpha_2$, $\eta_2$,
and a rotation angle $\beta_2$ deduced from $\alpha_1$, $\eta_1$,
$\beta_1$ by replacing $\omega_2$ by $\omega_3$.

\subsubsection{Potential energy}
In calculating the matrix elements of the potential operator, one can
focus on the operator (\ref{pot}) acting on the pair $kl$. Then
\begin{equation}
\langle \eta_{\sigma_i} \Phi_{k_i} \mid V_{kl} \mid \eta_{\sigma_j}
\Phi_{k_j} \rangle = \sum_s  {\cal O}_{ij}^{(s)}(kl)
{\cal E}_{ij}^{(s)}(kl),
\end{equation}
where
${\cal O}_{ij}^{(s)}(kl)$ is the matrix element of the internal
operator between the internal wave functions. It is calculated in
practice by Racah techniques. We are interested here by the matrix
element concerned with the space part
\begin{equation}
{\cal E}_{ij}^{(s)}(kl)=\langle \Phi_{k_i} \mid V_{kl}^{(s)}(\mid {\bf
r}_k -{\bf r}_l \mid) \mid \Phi_{k_j} \rangle.
\end{equation}

The term $V_{23}$ is easy to calculate since the argument entering
this term is precisely one of the Jacobi coordinate. One gets
\begin{equation}
\label{e23}
{\cal E}_{ij}^{(s)}(23)=\delta({\bar k}_i,{\bar k}_j)
V_{n_il_i,n_jl_j}^{(s)}(b_x)
\end{equation}
with the definition (\ref{potoh}) for the matrix element in the HO
basis.

The calculation for $V_{13}$ is more involved but can be performed
with the same trick as the one used for $K_2$. The argument
${\bf r}_{13}$ appearing in the potential can be put in the form
${\bf r}_{13} = \alpha_1 {\bf u}$ with the vector ${\bf u}$ obtained
from ${\bf x}$ and ${\bf y}$ by a rotation with the angle $\beta_1$.
The coefficients $\alpha_1$ and $\beta_1$ are already defined in
(\ref{para1}) and ({\ref{beta1}). One then introduces the vector
${\bf v}$
orthogonal to vector ${\bf u}$. The basis state are HO functions with
({\bf x},{\bf y}) representation; with appropriate BMC we change them
into HO functions with ({\bf u},{\bf v}) representation. In this
representation, the matrix element is obvious. The result is
\begin{eqnarray}
\label{e13}
{\cal E}_{ij}^{(s)}(13)
&=&\delta_{L_i L_j} \pi_2 \sum_
{n_2,n_2',l_2,\nu_2,\lambda_2} \langle \nu_2 \lambda_2 n_2 l_2; L_i
\mid n_j l_j \nu_j \lambda_j;L_i \rangle_{\beta_1} \nonumber \\
&& \langle \nu_2 \lambda_2 n_2' l_2; L_i \mid n_i l_i \nu_i \lambda_i;
L_i \rangle_{\beta_1} V_{n_2'l_2,n_2l_2}^{(s)}(\alpha_1).
\end{eqnarray}

The calculation for $V_{12}$ is performed in an analogous way; the
value of ${\cal E}_{ij}^{(s)}(12)$ has the same expression as
(\ref{e13}) but with a phase $\pi_3=1$ and with $\alpha_2$ and
$\beta_2$ replacing $\alpha_1$ and $\beta_1$.

\subsubsection{Differences with the usual method}

In this part, we want to point out what are the complications due to
the use of HO functions with different sizes as compared to the
traditional method based on HO functions with a unique size.

First, the number of basic states is a geometrical property based on
invariance principles acting on quantum numbers; thus, for a given
number of quanta $N_Q$, the number of basis states in both methods is
exactly the same. The diagonalization procedure takes more or less the
same time.

Now, one must examine the time needed to compute the matrix elements.
In fact, one should realize that their formal expressions are exactly
the same in both approaches. Thus, if we suppose that $b,b_x,b_y$ are
given once for all, the new method is as easy (or as difficult!) and
as fast as the old one.

The difference is merely in the determination of the size parameters.
In both method they are determined by requiring a minimum for the
energy of a particular state. In the old method we have only one
parameter $b$ whereas in this new method the minimization must be done
in a two dimensional space ($b_x,b_y$). This results of course in a
larger time. But there is also another complication which is less
transparent. In the old method the BMC to be used in the formalism
depend only on the $\omega_i$ parameters (on the system) but not on
the $b$ parameter; this was the reason for choosing the special set of
Jacobi coordinates (\ref{trdjc}). Thus they can be calculated once for
all at the beginning of the code and remain the same during the
variational procedure. In the new method the BMC depend both on
$\omega_i$ {\em and} ($b_x,b_y$) (see relations~(\ref{ksr2}) and
(\ref{beta1})) so that
they need to be recalculated at each step of the variational
procedure. At first sight this may seem a dramatic drawback; however
this must be moderated because BMC are computed very fast, and also
because, for the same precision, the matrices in the new method are
smaller than in the old one. All these aspects are commented later on.

\subsection{Identical particles}
\label{subsec:iden}

In the case of two identical particles, it is natural to consider
them as the objects 2 and 3, with the set of Jacobi coordinates chosen
(\ref{jc}). It is then easy to select, in all the possible basis
states,
those characterized by the good symmetry property. This implies some
constraints (depending on the fact that the particles are fermions or
bosons) on the quantum numbers of the wave functions associated with
the variable {\bf x}. The basis is then smaller (roughly by a factor
2)
 than in the case of
three different particles, and it can be also shown that
\begin{eqnarray}
\label{twoid}
\langle \Phi_{k_i} \mid K_{2} \mid \Phi_{k_j} \rangle &=&
\langle \Phi_{k_i} \mid K_{3} \mid \Phi_{k_j} \rangle, \nonumber \\
\langle \eta_{\sigma_i} \Phi_{k_i} \mid V_{12} \mid \eta_{\sigma_j}
\Phi_{k_j} \rangle &=& \langle \eta_{\sigma_i} \Phi_{k_i} \mid V_{13}
\mid \eta_{\sigma_j} \Phi_{k_j}\rangle.
\end{eqnarray}
Consequently, the computation labor is in this case greatly reduced.

When the three particles are identical, it is not obvious to build the
basis states in such a way that they are all completely symmetrical or
antisymmetrical for the permutation of the particles. If we
diagonalize the Hamiltonian in the basis for which particles 2 and 3
have already good symmetry properties, we obtain eigenstates which are
either
completely
symmetrical, completely antisymmetrical, or of mixed
symmetry. A way to distinguish all these states is to calculate for
each state the mean value of the transposition operator $P_{13}$ for
particles 1 and 3. The completely symmetrical (antisymmetrical) states
will be characterized by $\langle P_{13} \rangle = +1$ ($-1$).
One can thus imagine to let the Hamiltonian do the job to filter
states with given symmetry, verify a posteriori the symmetry of
eigenstates,
and reject those having a symmetry not compatible with the Pauli
principle.
Practically this procedure cannot be applied systematically
because very often there exist degenerate states with different
symmetries.

This is why we adopt an approach which is more painful but which
works correctly each time. In practice, we diagonalize the operator
$P_{13}$ in the basis
symmetrized for particles 2 and 3. We select the eigenstates with
eigenvalues $+1$ or $-1$ according to the nature of our particles.
Then
we diagonalize the Hamiltonian in the basis built with the selected
eigenstates. We can also perform the inverse basis change to obtain
the Hamiltonian eigenstates expressed in the original basis.

We have to compute the matrix elements
$\langle \eta_{\sigma_i} \Phi_{k_i} \mid P_{13} \mid \eta_{\sigma_j}
\Phi_{k_j} \rangle$. The mean value of the operator for color, isospin
and spin degrees of freedom is very easy to calculate by usual Racah
techniques. The computation for the space part is much more involved.
Let us note (${\bf x}',{\bf y}'$) the coordinates resulting of the
action of $P_{13}$ on the coordinates (${\bf x},{\bf y}$). Then we
have
\begin{equation}
\label{meanp13}
\langle \Phi_{k_i} ({\bf x},{\bf y}) \mid P_{13} \mid \Phi_{k_j} ({\bf
x},{\bf y}) \rangle = \langle \Phi_{k_i} ({\bf x},{\bf y}) \mid \Phi_{
k_j} ({\bf x}',{\bf y}') \rangle.
\end{equation}
The trick is to introduce new sets of coordinates
(${\bf u},{\bf v}$)
and rotations $\cal R$ of angle $\varphi$ such that, for instance,
\begin{equation}
\label{rotuv}
\left( \begin{array}{c} {\bf x}' \\ {\bf y}' \end{array} \right) =
{\cal R}
\left( \begin{array}{c} {\bf u} \\ {\bf v} \end{array} \right)
\quad {\rm and} \quad
\left( \begin{array}{c} {\bf x} \\ {\bf y} \end{array} \right) =
{\cal R}
\left( \begin{array}{c} {\bf v}/\sqrt{\alpha} \\ \sqrt{\alpha}\,
{\bf u}
\end{array} \right).
\end{equation}
It is then possible to calculate the matrix element (\ref{meanp13})
\begin{eqnarray}
\label{p13gen}
\langle \Phi_{k_i} \mid P_{13} \mid \Phi_{k_j} \rangle &=&
\delta_{L_iL_j}
\ \ \pi (r)
\sum_{n_1,n_1',l_1,\nu_1,\nu_1',\lambda_1}
\langle \nu_1' \lambda_1 n_1' l_1; L_i
\mid n_j l_j \nu_j \lambda_j;L_i \rangle_{\varphi}\nonumber \\
&&\langle n_1 l_1 \nu_1 \lambda_1 ; L_i
\mid n_i l_i \nu_i \lambda_i; L_i \rangle_{\varphi}\:
F_{n_1' n_1 l_1}(\sqrt{\alpha})\: F_{\nu_1' \nu_1 \lambda_1}(1/\sqrt{
\alpha}),
\end{eqnarray}
where
\begin{eqnarray}
\label{p13gen2}
&&\alpha= \frac{16 r^4+8r^2+9+\rho}{32r^2}, \quad
\rho=\sqrt{(4r^2-3)^2(16r^4+40r^2+9)}, \quad r=\frac{b_y}{b_x},
\nonumber \\
&&\cos\varphi=\sqrt{\frac{\rho-16r^4+9}{2\rho}}, \quad
\sin\varphi=\sqrt{\frac{\rho+16r^4-9}{2\rho}}, \quad {\rm and}
\nonumber \\
&&\pi (r)=(-1)^{\lambda_i+l_i+L_i} \quad {\rm if}\quad r < \sqrt{3}/2
\quad {\rm or}
\quad \pi (r)=(-1)^{\lambda_i+l_j+L_i} \quad {\rm if}\quad r >
\sqrt{3}/2.
\end{eqnarray}
The quantity $F_{n' n l}(a)$ measures the overlap of one HO function
with
another one scaled by a positive factor
\begin{equation}
\label{fnnl}
F_{n' n l}(a) = a^{3/2} \int_0^{\infty} u_{n'l}(x)\: u_{nl}(ax)\: dx.
\end{equation}
Its analytical expression as well as its symmetry properties are
given in Ref.~\cite{sema95}.

From Eq.~(\ref{p13gen}), it appears that the operator $P_{13}$
couples basis states which can be characterized by different numbers
of
quanta. Consequently, in a basis truncated at a fixed number of
quanta, it is not possible to obtain an integer value for
$\langle P_{13} \rangle$, that is to say an eigenstate with a well
defined symmetry. Such an eigenstate needs an infinite number of
basis states to develop.

Nevertheless, if $b_y/b_x = \sqrt{3}/2$, that is to say if the
relationship (\ref{lien}) is verified, we have $\alpha = 1$
and $\varphi = \pi/6$, which implies
\begin{equation}
\label{p13}
\langle \Phi_{k_i} \mid P_{13} \mid \Phi_{k_j} \rangle =
\delta_{L_iL_j}
(-1)^{\lambda_i+l_i+L_i}
\langle \nu_j \lambda_j n_j l_j ; L_i
\mid n_i l_i \nu_i \lambda_i ;L_i \rangle_{\pi/6}.
\end{equation}
In this case, two basis states with different numbers of quanta are
not mixed by the operator $P_{13}$, and it is possible to obtain an
eigenstate with a defined symmetry in a basis truncated at a fixed
number of quanta. This is the reason why we work in such bases, as
mentioned above.

To study systems with three identical particles we must choose between
two procedures. We can work with $b_x$ and $b_y$ completely free to
compute the lowest possible upper bounds, but the price to pay is to
obtain eigenstates which are not characterized by a defined symmetry.
On contrary, we can impose the constraint (\ref{lien}) on $b_x$ and
$b_y$ to get eigenstates with a defined symmetry, but with the risk
to not obtain the lowest possible upper bounds. In all cases studied,
we remarked that it is preferable to work with the second procedure
because the loss of good symmetry properties results in an increase of
the upper bounds which cannot be compensated by relaxing the
constraint (\ref{lien}). Actually, the asymmetry between Jacobi
coordinates is less pronounced in three identical particle
systems, it is then not a serious penalty to work with only one
effective variational parameter for such systems.

\section{Numerical aspects}
\label{sec:num}

This section is devoted to some tricks that we employed in our
numerical codes to fasten the computations.

As we saw just before, the BMC need to be calculated very often, each
time as we change one of the size parameters. Moreover, if the number
of quanta increases, the number of BMC required increases also
drastically. The algorithm to calculate them has been explained in
detail in Ref.~\cite{sil85}; it relies on recursive formulae which are
precise and fast enough. In order to be efficient this algorithm needs
to calculate all of them up to a given number of quanta $N_Q$ even if
they are not all necessary for our calculations. Table \ref{tab:bmc}
shows the total number of BMC as a function of $N_Q$. In our
calculations
we have pushed the expansion up to $N_Q=16$. The great advantage of
this algorithm is that the BMC are stored naturally in such a way that
the elements needed in the various summations where they appear
(summations over $n,l,\nu,\lambda$) are placed contiguously in a one
dimensional array so that the summation is restricted to a reading in
sequence which is very fast.

Another time consuming part of our job, is the calculation of the
matrix elements (\ref{potoh}) which appear in the inner loops of our
codes. There exists a very old way, that is also often forgotten, to
calculate them precisely and very fast. The technique relies on
Talmi's integrals and was reported elsewhere \cite{sema95}. More
details are
provided in the appendix~\ref{sec:talmi}. Let us just mention that
most of Talmi's
integral of practical use can be evaluated analytically; this is
important because they depend on the size parameters and must be
calculated very often.

\section{Results}
\label{sec:res}

Our method can be applied to a wide variety of three particle systems.
In this paper we study the convergence rate with baryons considered as
three quark systems. We report some results obtained with a
nonrelativistic potential model which can describe quite well meson
and baryon spectra \cite{bhad81}, and two simple potential models
developed to compare nonrelativistic and
semirelativistic approaches \cite{fulc94}.

The quality of an upper bound depends on two kinds of parameters: the
number of quanta and the oscillator length parameters. In the case of
three identical particles, we have mentioned that it is preferable to
work with an unique parameter $b$ (see Sec.~\ref{subsec:iden}). In
Fig.~\ref{fig:eb}, the nucleon mass $M_N$ for the model of
Ref.~\cite{bhad81} is plotted as a function of $b$ for different
values
of the number of quanta $N_Q$. For small values of $N_Q$, a good
choice of $b$ is crucial, but as $N_Q$ increases, the minimum of the
curve $M_N(b)$ becomes more and more flat. In order to save
computation time, it is interesting to compute a given upper bound in
two steps. First, determine the optimum value of the oscillator
parameter for this upper bound computed with a small value of the
number of quanta, say $N'_Q$. Secondly, use this value of the
oscillator
parameter to recompute the upper bound with a higher value of the
number of quanta, say $N_Q$. This situation is illustrated in
Table~\ref{tab:qqr} for two $I=S=1/2$ lowest state baryons within the
model of Ref.~\cite{bhad81}. We can see that for baryon containing at
least two different particles, it is also interesting to use this
procedure. In the following we will always take $N'_Q=8$. The maximum
number of quanta used in this paper is $N_Q=16$. A higher value is not
considered for practical reasons (see Table~\ref{tab:bmc} and appendix
~\ref{sec:talmi}). The method used to determine the optimum values of
length parameters is described in appendix~\ref{sec:mini}.

We will now see that good upper bounds can be obtained with the
procedure described above. First look at the case of 3 identical
particles, for which it is preferable to use the constraint
(\ref{lien}) on length parameters. In Table~\ref{tab:nrsr}, binding
energies of
the center of gravity $N-\Delta$ for the nonrelativistic and
semirelativistic Fulcher's models \cite{fulc94} are given as a
function of the number of quanta $N_Q$. For both kinematics, the
convergence is reached at $N_Q=16$.

In the cases of asymmetric systems, we can expect that the use two
nonlinear parameters will bring some advantages. In
Table~\ref{tab:fulcnr}, binding energies of the lowest state $ubb$
baryon, for the
nonrelativistic Fulcher's models \cite{fulc94}, are given as a
function of the number of quanta $N_Q$, for two values of $L$. For
the $L=0$ state, the use of two oscillator lengths yields only a very
small improvement for a small number of quanta. This improvement even
vanishes when the number of quanta increases. For $L=4$, the upper
bound is significantly below when two nonlinear parameters are used at
small number of quanta. The result at $N_Q=8$ with two parameters is
better than the one at $N_Q=16$ with only one parameter. This means
that if a
tremendous precision is not necessary one can be content with a small
value of $N_Q$ for two oscillator lengths. This implies, for instance,
work with 50 basis states instead of 420 (see Table~\ref{tab:fulcnr}
for $L=4$). Similar results are obtained with the nonrelativistic
model of Ref.~\cite{bhad81}.

For semirelativistic kinematics, the new method yields more drastic
improvement. In Table~\ref{tab:fulcsr}, binding energies of the lowest
state $ubb$ baryon,
for the semirelativistic Fulcher's models \cite{fulc94}, are given as
a function of the number of quanta $N_Q$, for two values of $L$. As
well for $L=0$ as for $L=4$, the upper bounds at $N_Q=8$ with two
nonlinear parameters is much better than the ones at $N_Q=16$ with one
nonlinear parameter, the gain being larger for $L=4$. Moreover with
only one oscillator length, the convergence is not reached, contrary
to the situation with two oscillator lengths. Again, if a great
precision is not crucial, one can be content with small value of $N_Q$
for two oscillator lengths.

It is worth noting that in the case of three different particles, a
good choice of the numbering of particles can increase the
convergence rate. In Table~\ref{tab:ucb}, the binding energy of the
lowest state of the $ucb$ $L=4$
baryon for the semirelativistic Fulcher's models \cite{fulc94}
is computed as a function of the number of quanta $N_Q$. For small
values of $N_Q$, we can see that the coupling $u[cb]$ gives lower
bounds that the coupling $c[ub]$. With the first numbering, we
benefit at best of the asymmetry of the system: $c$ and $b$ quarks
form a small diquark with the $u$ quark orbiting around. The
oscillator length $b_x$, associated with the coordinate
${\bf r}_2-{\bf r}_3$, is smaller than the parameter $b_y$. The
situation is at the opposite for the coupling $c[ub]$. The difference
seems small but it is large enough to give lower bounds. Obviously
when the number of quanta increases, both coupling methods tend to
give the same results, since the mass of real state (infinite number
of quanta) is independent of the numbering of the particles.

Our method is mainly efficient in the case of semirelativistic
kinematics. One can ask if it is really important for three-body
systems. Several works have shown that semirelativistic
kinematics is a key
ingredient of quark potential models (see for instance \cite{isgur}).
Here, we illustrate this point with simple calculations relying on
potential models used above. In Ref.~\cite{fulc94}, it is shown that
a semirelativistic potential model yields a better description of
meson
spectra than a nonrelativistic approach. We will use the two models of
this paper to compute some baryon masses in order to see if the
semirelativistic kinematics is again preferable.

Hamiltonian described in Ref.~\cite{fulc94} do not contain any spin
nor isospin dependent operator, so it is only possible to compute
center of gravity of families of baryon. In Table~\ref{tab:ground},
some ground states and first excited states of strange and non-strange
baryons are compared with experimental data. For both nonrelativistic
and semirelativistic spectra a simple three-body term has been added
in order to obtain exactly the $N-\Delta$ center of gravity. This
term, proposed in Ref.~\cite{bhad81}, is a constant $A$ divided by the
product of the three quark masses contained in the baryon. The
``experimental'' centers of gravity are obtained on the basis of a
chromomagnetic description of baryons (see for instance
\cite[p.~384]{clos79}). A $\chi^2$ value is computed for each
kinematics, with a standard deviation estimated at 15 MeV, around the
isospin breaking value. We can clearly see that the semirelativistic
approach is far better, due mainly to a much more reasonable
description of first excited states.

Orbital excitations of baryons are also better described by
relativistic kinematics. This can be seen on Fig.~\ref{fig:regge},
where the predictions of the two models of Ref.~\cite{fulc94} for the
Regge trajectory of the $\Delta$-family are plotted. For this figure
both spectra are renormalized in order to give the exact mass for the
baryon $\Delta$.

It is worth mentioning a phenomenon which can complicate the search of
an optimal upper bound. On Fig.~\ref{fig:cross}, the five first
binding energies of a very asymmetric baryon are plotted as a function
of an unique oscillator length $b$. At a first glance, one can see
several crossings of levels for particular value of $b$. If we zoom on
these points, we can see that there is no crossing at all actually. We
have remarked that the apparition of (what we call) pseudo-crossing is
favoured for very asymmetric systems, high angular momenta and
semirelativistic kinematics. It is worth noting that the value of $b$,
for which a pseudo-crossing between two given levels appears,
decreases when the number $N_Q$ of quanta increases. Indeed, when
$N_Q$ increases, a wider range of values of $b$ allows to obtain a
good approximation of the wave functions; the unphysical
characteristics of the spectrum are rejected toward zero length
parameter. Similar phenomena appears when two oscillator lengths are
considered, but they are much more difficult to visualize. Sometimes,
the
presence of pseudo-crossings can perturb the search of a minimum since
the binding energy can vary abruptly with the length parameter at
these points. The solution is simply to take smaller ranges of $b$
values to search for the minimum energy.

\section{Conclusions}
\label{sec:conclu}

The quantum three-body problem is well under control from the
numerical point of view. In this paper we have revisited the method
based on an expansion of the wave function on HO basis. The
advantage of this approach is the possibility to allow different sizes
$b_x$ and $b_y$ for HO functions related to different Jacobi
coordinates ${\bf x}$ and ${\bf y}$.

We proved that the matrix elements can be calculated without any
approximation and exactly for any value of the number of quanta $N_Q$.
The complications as compared to the traditional approach is that we
are obliged now to perform a double minimization on $b_x$ and $b_y$
instead of a single minimization on an unique parameter $b$; moreover
the Brody-Moshinsky coefficients need also to be recalculated each
time we change the size parameters. These disagreements are largely
compensated by the fact that, for a given precision, the matrices to
be diagonalized are much smaller. This last point is equivalent to say
that for a given number of quanta, the precision achieved can be
largely increased as compared to the old method. We thus think that
our new treatment is a good compromise between precision and numerical
effort.

Since it is universal and systematic, our method is particularly well
suited for very asymmetric systems (for example one light and two
heavy particles) and for systems having a large orbital angular
momentum ($L$ = 4, 5, 6, \ldots). This is not the case for most of
other competitive approaches. Moreover, the method works particularly
well for semirelativistic kinematics. We have been very careful to
include several options (storage of BMC, use of Talmi's integrals,
special minimization procedure, \ldots) that allow a drastic gain in
computer time. The numerical code can be adapted on any normal
personal computer and the results are already very satisfactory even
with a few tens of seconds run on these machines.

Besides the numerical aspect, which is however very important, this
method deals with more appropriate degrees of freedom and thus sticks
more to the physical system. In particular a good precision can be
achieved with a wave function containing a quite reasonable number of
basic states $N$. This point is very important for the calculation of
physical observables which grows as $N^2$.

For the moment our code can deal with nonrelativistic and
semirelativistic kinetic energy terms and with central and hyperfine
potentials. It can be adapted, with some modifications, to treat also
more complex structures such as instanton effects, spin-orbit and
tensor forces. The treatment of three-body interaction can also be
considered. Some of these aspects are already under work.

\acknowledgments

C. Semay would like to thank the F.N.R.S. for financial support, and
F. Brau would like to thank the I.I.S.N. for financial support.

\appendix

\section{Talmi's integrals}
\label{sec:talmi}

The computation of matrix elements (\ref{potoh}) can be performed in a
very efficient way by means of the so-called Talmi's integrals. It
can be shown that \cite{bro}
\begin{equation}
\label{emho}
V_{nl,n'l'}(b) = \delta_{ll'}\sum_{p=l}^{n+n'+l}
B(n,n^{\prime},l,p)\: I_p(V,b).
\end{equation}
In presence of
tensor forces, formula~(\ref{emho}) must be modified as states with
different orbital angular momenta are mixed. In this case, new
coefficients $B(n,l,n',l',p)$ must be used.
In expression~(\ref{emho}), the quantities $B(n,n^{\prime},l,p)$ are
geometrical coefficients which can be calculated once for all, while
the numbers $I_p(V,b)$ are the Talmi's integrals which must be
computed each time the length scales of the HO functions are changed.
They are explicitly given by the following formula
\begin{equation}
\label{intl}
I_p(V,b) = \frac{2}{\Gamma(p+3/2)}\int_0^\infty x^{2p+2}\:\exp(-x^2)\:
V(bx)\:dx.
\end{equation}
This method has two great advantages: i) only $2N+1$ Talmi's integrals
are necessary to get the $N(N+1)/2$ matrix elements, which save a
lot of computation time; ii) most of the $I_p(V,b)$ quantities are
given by an analytical expression, so that the complete set of matrix
elements are obtained fast and with a good precision.

It is worth mentioning that the coefficients $B(n,n',l,p)$ can be
stored in such a way that the elements needed in a summation where
they appear are placed contiguously in a one dimensional array so that
the summation is restricted to a reading in sequence which is very
fast. These coefficients can be computed very accurately but their
values increase rapidly with the quantum numbers. As the
summation~(\ref{emho}) is an alternate one, the values of the Talmi's
integrals result from differences of large numbers. Working with
double precision numbers limits the use of this technique for values
of $n+n'+l$ below a number around 20.

Some Talmi's integrals for various potentials are given in
Ref.~\cite{sema97}. For the nonrelativistic kinetic energy term, it
is not necessary to use Talmi's integral since the matrix elements of
{\bf p}$^{2}$ on HO functions are very simple expressions
(see formula~(\ref{kp})). The analytical form for the matrix elements
of the semirelativistic kinetic energy operator (\ref{calk}) involves
the calculation of the following Talmi's integral
\begin{equation}
\label{talsr}
I_p \left( e^{-ax^2}\sqrt{b x^2+c} \right)= \sqrt{b}\left(\frac{c}{b}
\right)^{p+2} U
\left(p+\frac{3}{2}, p+3, \frac{(1+a)c}{b}\right).
\end{equation}
$U(x,y,z)$ is a Kummer function \cite{abra70}, which can be
calculated accurately by using recurrence formula for small values of
$z$, or asymptotic expansion for large values of $z$. For medium
values
of this parameter, a direct integration of (\ref{intl}), by
Gauss-Laguerre method for instance, gives the best accuracy.

\section{Minimization procedure}
\label{sec:mini}

For a given number of quanta, the quality of the lower bound $E_k$ for
the $k$th level depends on the length scales parameters $b_x$ and
$b_y$. As we have seen in Sec.~\ref{subsec:iden}, only one parameter
$b$ is relevant in the case of three identical parameters. It is then
necessary to find a fast method to compute the minimum of the
functions $E_k(b_x,b_y)$ or $E_k(b)$. Let us focus first on the case
of one nonlinear parameter.

A very efficient algorithm to find the minimum of a one parameter
function is the Brent's method \cite{pres92}. It relies on successive
approximations of the function by parabolic curves. This method is
robust and necessitates only the computation of one new point at each
iteration, but it presents 3 drawbacks: i) to start, three abscissas
must be given in a such way that the second one corresponds to the
lowest ordinate; ii) nothing prevents the algorithm to find a new
abscissa with a value irrelevant for the problem chosen; iii) the real
form of the function can be very different of a parabola in the first
steps of the procedure, which can increases dramatically the
computation time. A way to cure simultaneously these flaws is to
approach the function to minimize, at least for the first iterations,
by a trial function presenting one minimum and which matches at best
the real functions in the relevant range of abscissa values. In order
to not penalize the method, the trial function must be defined with
only three parameters as a parabola. One can try
\begin{equation}
\label{formin}
y = \alpha x^m + \beta x^n + \gamma x^p,
\end{equation}
where $m$, $n$ and $p$ are different fixed real numbers. Given 3
values
$y_1$, $y_2$ and $y_3$, for 3 given values $x_1$, $x_2$ and $x_3$, the
parameters $\alpha$, $\beta$ and $\gamma$
can be found analytically if one power vanishes or if $n=(m+p)/2$. So
we always work with these constraints.

Applied to the study of baryons with nonrelativistic kinematics for
instance, we found that the choice $m=-2$, $n=1$ and $p=0$ allows a
fast computation of the minimum of the curve $E_k(b)$, even if the
position of the minimum is very badly estimated. These numbers stems
from the dependence on $b$ of the Talmi's integral. Other sets of
numbers can be easily found in the case of different interactions.

To search the minimum of the functions $E_k(b_x,b_y)$, we apply our
modified Brent's method alternatively for parameters $b_x$ and $b_y$.
With a judicious management of the search procedure, this method is in
most cases faster and safer than more sophisticated algorithms.

\begin{table}
\protect\caption{Number of Brody-Moshinsky coefficients (BMC) to be
calculated for a given number of quanta $N_Q$. For details concerning
the algorithm, see for example Ref.~\protect\cite{sil85}.}
\label{tab:bmc}
\begin{tabular}{rrrrr}
\hline
$N_Q$ & Number of BMC & $N_Q$ & Number of BMC \\
\hline
0 & 1 \\
1 & 5 & 9 & 12~225 \\
2 & 24 & 10 & 22~352  \\
3 & 80 & 11 & 39~136  \\
4 & 240 & 12 & 66~168  \\
5 & 616 & 13 & 108~264  \\
6 & 1~456 & 14 & 172~320  \\
7 & 3~144 & 15 & 267~312  \\
8 & 6~389 & 16 & 405~537  \\
\hline
\end{tabular}
\end{table}

\begin{table}
\protect\caption{Binding energies $E$ (in MeV) of two lowest state
$I=S=1/2$ baryons for the potential of
Bhaduri {\em et al.} \protect\cite{bhad81}. Energies are calculated
with
16 quanta for values of oscillator parameters (in GeV$^{-1}$) which
give the lowest energy for $N'_Q$ quanta.}
\label{tab:qqr}
\begin{tabular}{rrrrrr}
\hline
\ & \multicolumn{2}{c}{$uuu$, $L=0$} &
\multicolumn{3}{c}{$ubb$, $L=4$} \\
$N'_Q$ & $b$ & $E$ & $b_x$ & $b_y$ & $E$ \\
\hline
0 & 3.337 & 17.956 & \ & \ & \ \\
2 & 3.238 & 17.466 & \ & \ & \ \\
4 & 2.941 & 16.167 & 1.739 & 2.711 & 55.558  \\
6 & 2.817 & 15.695 & 1.759 & 2.668 & 55.535  \\
8 & 2.648 & 15.117 & 1.704 & 2.474 & 55.422  \\
10 & 2.551 & 14.822 & 1.719 & 2.348 & 55.368  \\
12 & 2.425 & 14.495 & 1.654 & 2.236 & 55.319  \\
14 & 2.347 & 14.345 & 1.699 & 2.110 & 55.290  \\
16 & 2.251 & 14.262 & 1.619 & 2.041 & 55.278  \\
\hline

\end{tabular}
\end{table}

\begin{table}
\protect\caption{Binding energies (in MeV) of the center of gravity
$N-\Delta$
for the nonrelativistic (NR) and semirelativistic (SR) Fulcher's
models \protect\cite{fulc94} as a function of the number of quanta
$N_Q$. Energies are computed with the value of the oscillator
parameter $b$ (in GeV$^{-1}$) which
gives the lowest bound for $8$ quanta. The dimension $D_Q$ of the
basis for $N_Q$ quanta is also indicated. Note that values of the
binding energies are very different for the two models since the mass
of quark $u$ is 0.325 GeV for the NR model and 0.150 GeV for the SR
model.}
\label{tab:nrsr}
\begin{tabular}{rrrr}
\hline
$N_Q$ & $D_Q$ & NR & SR \\
\ & & $b=3.119$ & $b=2.639$ \\
\hline
8 & 70 & 244.35 & 799.85 \\
10 & 112 & 244.28 & 799.56 \\
12 & 168 & 244.17 & 798.61 \\
14 & 240 & 244.14 & 798.45 \\
16 & 330 & 244.11 & 798.11 \\
\hline
\end{tabular}
\end{table}

\begin{table}
\protect\caption{Binding energies (in GeV) of the lowest state $S=1/2$
$ubb$ baryon, for two values
of $L$, for the nonrelativistic Fulcher's models
\protect\cite{fulc94}, as a function of the number of quanta $N_Q$.
Energies $E_1$ are computed with an unique value of the oscillator
parameter $b$, while energies $E_2$ are computed with $b_x$ and $b_y$
not constrained by the relation (\ref{lien}). Energies are computed
with values of the oscillator parameters (in GeV$^{-1}$) which give
the lowest bound for $8$ quanta. The dimension $D_Q$ of the
basis for $N_Q$ quanta is also indicated.}
\label{tab:fulcnr}
\begin{tabular}{cccc}
\hline
$N_Q$ & $D_Q$ & $E_1$     & $E_2$ \\
 \hline
$L=0$ & & $b=3.592$ & $b_x=0.996$ \\
      & &           & $b_y=2.260$ \\
\hline
 8 & 35 &  0.3664  & 0.3661  \\
10 & 56 &  0.3661  & 0.3659  \\
12 & 84 &  0.3654  & 0.3654  \\
14 & 120 & 0.3652  & 0.3652  \\
16 & 165 & 0.3650  & 0.3650  \\
\hline
$L=4$ & & $b=5.521$ & $b_x=1.766$ \\
      & &           & $b_y=2.565$ \\
\hline
 8 & 50 &  1.0929  & 1.0703  \\
10 & 100 & 1.0783  & 1.0699  \\
12 & 175 & 1.0732  & 1.0696  \\
14 & 280 & 1.0713  & 1.0695  \\
16 & 420 & 1.0705  & 1.0694  \\
\hline
\end{tabular}
\end{table}

\begin{table}
\protect\caption{Same as for Table~\ref{tab:fulcnr} but for the
semirelativistic Fulcher's models
\protect\cite{fulc94}.}
\label{tab:fulcsr}
\begin{tabular}{cccc}
\hline
$N_Q$ & $D_Q$ & $M_1$     & $M_2$ \\
 \hline
$L=0$  & & $b=3.951$ & $b_x=0.963$ \\
       & &           & $b_y=1.650$ \\
\hline
 8 & 35 &  0.6242 & 0.5969 \\
10 & 56 &  0.6141 & 0.5966 \\
12 & 84 &  0.6067 & 0.5948 \\
14 & 120 & 0.6034 & 0.5945 \\
16 & 165 & 0.6003 & 0.5938 \\
\hline
$L=4$ & & $b=6.667$ & $b_x=1.651$ \\
      & &           & $b_y=2.147$ \\
\hline
 8 & 50 &  1.5441 & 1.3603 \\
10 & 100 & 1.4729 & 1.3591 \\
12 & 175 & 1.4304 & 1.3578 \\
14 & 280 & 1.4045 & 1.3573 \\
16 & 420 & 1.3885 & 1.3569 \\
\hline
\end{tabular}
\end{table}

\begin{table}
\protect\caption{Binding energies $E$ (in GeV) of the lowest state
$L=4$ $S=1/2$ $ucb$ baryon for the semirelativistic Fulcher's models
\protect\cite{fulc94}, as a function of the number of quanta $N_Q$,
and as a function of the coupling mode between particles.
Energies are computed with $b_x$ and $b_y$
optimum for $N'_Q$ quanta. The dimension $D_Q$ of the
basis for $N_Q$ quanta is also indicated.}
\label{tab:ucb}
\begin{tabular}{ccccccccc}
\hline
 & & \multicolumn{3}{c}{$c[ub]$} & \multicolumn{3}{c}{$u[cb]$} \\
$N'_Q$ & $N_Q$ & $D_Q$ & $b_x$ & $b_y$ & $E$ & $b_x$ & $b_y$ & $E$ \\
4 & 4 & 10 & 2.635 & 2.112 &  1.7394 & 2.131 & 2.442 & 1.6915 \\
6 & 6 & 40 & 2.646 & 2.087 &  1.6833 & 2.115 & 2.469 & 1.6718 \\
8 & 8 & 100 & 2.553 & 2.064 & 1.6638 & 2.091 & 2.343 & 1.6587 \\
8 & 10 & 200 & & &            1.6587 & & &             1.6566 \\
8 & 12 & 350 & & &            1.6562 & & &             1.6550 \\
8 & 14 & 560 & & &            1.6553 & & &             1.6545 \\
8 & 16 & 840 & & &            1.6546 & & &             1.6541 \\
\hline
\end{tabular}
\end{table}

\begin{table}
\protect\caption{Masses (in GeV) of some $L=0$
baryons for the nonrelativistic (NR) and semirelativistic (SR)
Fulcher's models \protect\cite{fulc94}, compared with experiment
(Exp.). All values given are masses relieved of chromomagnetic
contribution (see text). The
value of the corrective term is given for both model, as well
as the corresponding $\chi^2$ values.}
\label{tab:ground}
\begin{tabular}{rccc}
\hline
Baryons & Exp. & NR & SR \\
\hline
$N-\Delta$ ($1S$) & 1.086 & 1.086 & 1.086 \\
\phantom{$N-\Delta$} ($2S$) & 1.520 & 1.827 & 1.655 \\
$\Lambda-\Sigma-\Sigma^*$ ($1S$) & 1.269 & 1.250 & 1.269 \\
\phantom{$\Lambda-\Sigma-\Sigma^*$} ($2S$) & 1.735 & 1.928 & 1.832 \\
$\Xi-\Xi^*$ ($1S$) & 1.439 & 1.437 & 1.455 \\
$\Omega$ ($1S$) & 1.611 & 1.660 & 1.679 \\
\hline
$\displaystyle{\frac{A}{m_1m_2m_3}}$ (GeV) & & $-0.133$ & $-0.162$ \\
$\chi^2$ & & $119.4$ & $28.5$ \\
\hline
\end{tabular}
\end{table}

\begin{figure}
\protect\caption{Binding energy of the nucleon $E_N$ for the potential
of Bhaduri {\em
et al.} \protect\cite{bhad81} as a function of the number of quanta
$N_Q$ and the unique oscillator length $b$.}
\label{fig:eb}
\includegraphics*[width=10cm]{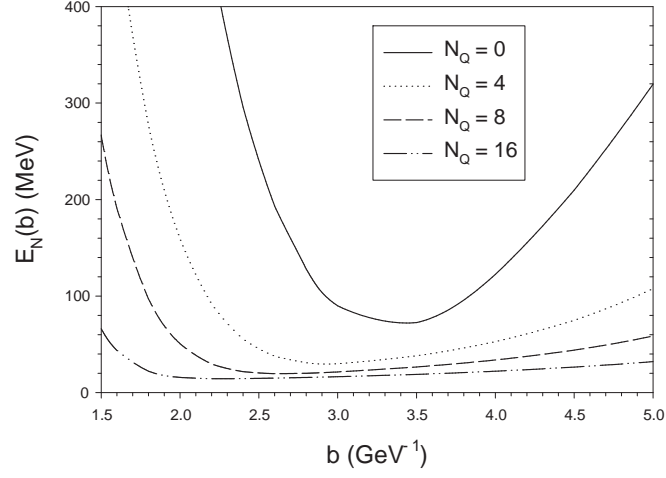}
\end{figure}

\begin{figure}
\protect\caption{Regge trajectories for the positive parity $I=3/2$
baryons. Circle indicates the experimental values with the error bars.
Solid (dashed) line joins the theoretical values from the
semirelativistic (nonrelativistic) Fulcher's model
\protect\cite{fulc94}.}
\label{fig:regge}
\includegraphics*[width=10cm]{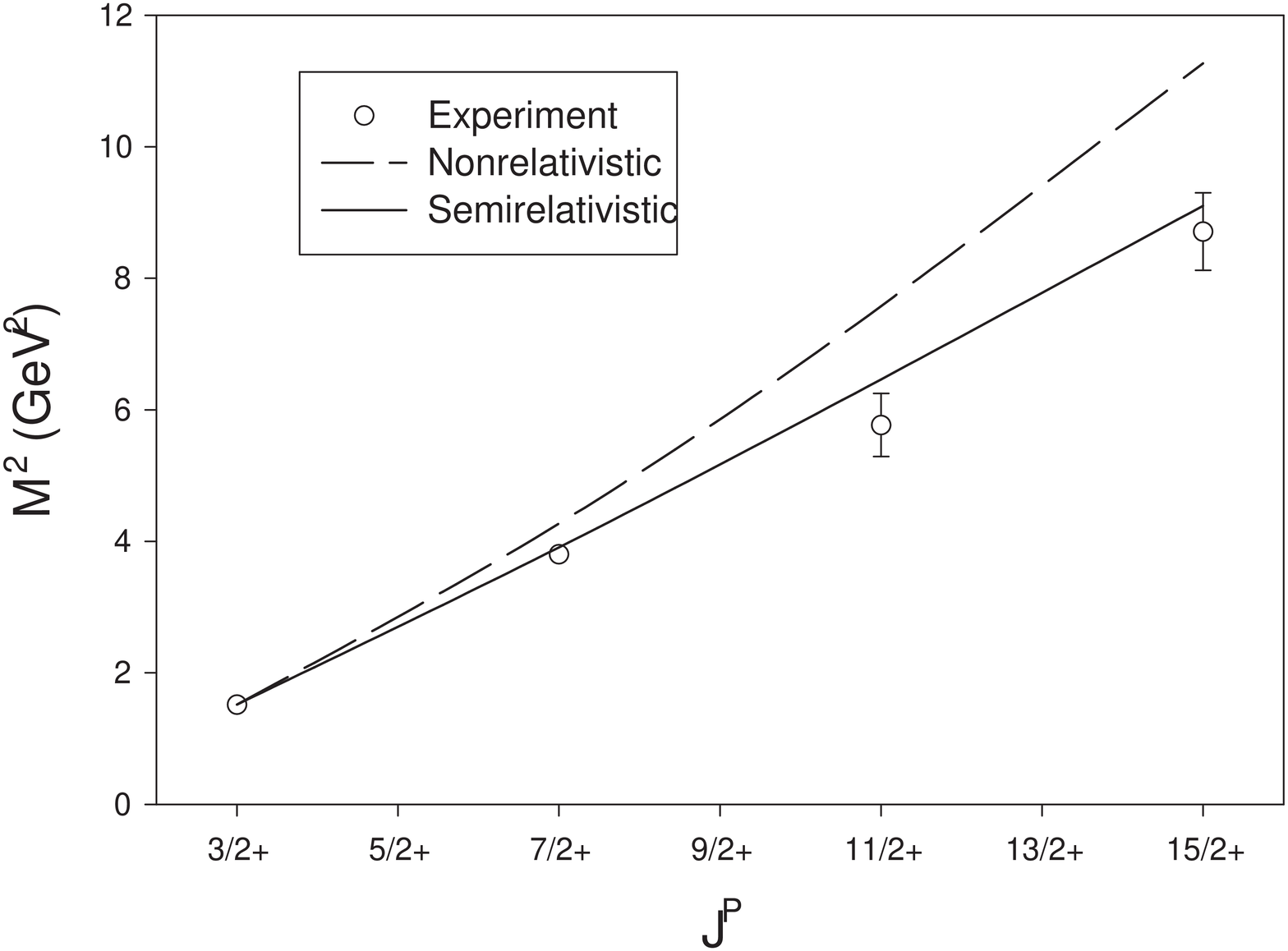}
\end{figure}

\begin{figure}
\protect\caption{Binding energies of the five first $L=4$ $S=1/2$
$ubb$
baryon states for the potential of Bhaduri {\em
et al.} \protect\cite{bhad81}, as a function of the unique oscillator
length $b$, for a number of quanta $N_Q=8$.}
\label{fig:cross}
\includegraphics*[width=10cm]{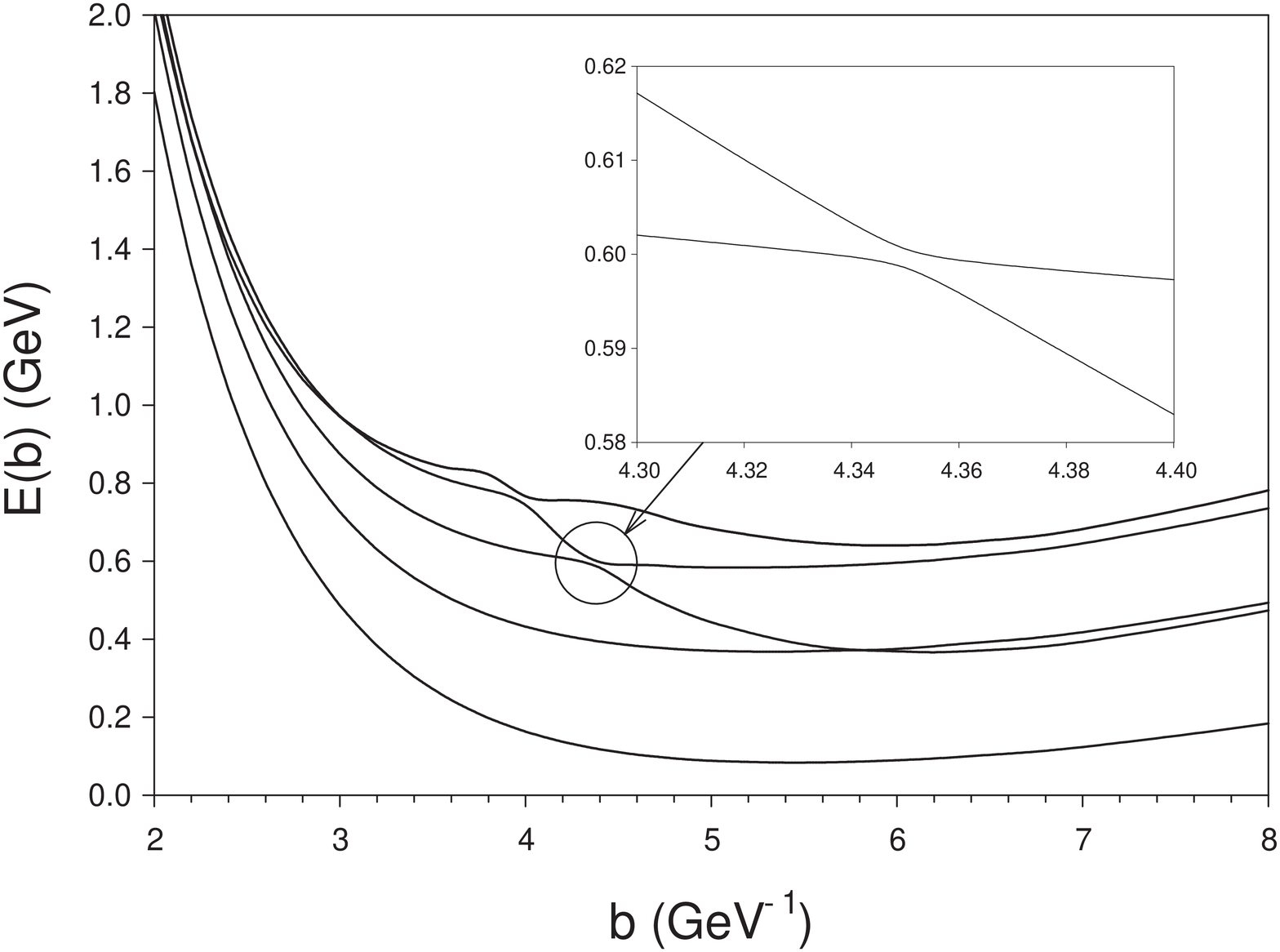}
\end{figure}


\begin{thebibliography}{99}

\bibitem{carl} J. Carlson, Phys. Rev. C {\bf 36}, 2026 (1987).
\bibitem{lind} W. von der Linden, Phys. Rep. {\bf 220}, 53 (1992).
\bibitem{ham} B. L. Hammond {\em et al.}, {\em Monte Carlo methods in
Ab initio quantum chemestry} (World Scientific, Singapore, 1994).
\bibitem{gloc} W. Gl\"{o}ckle, {\em The quantum mechanical few-body
systems} (Springer Verlag, Berlin, Heidelberg, 1983).
\bibitem{flr} M. Fabre de la Ripelle, Ann. Phys. {\bf 147}, 281
(1983).
\bibitem{suz} Y. Suzuki and K. Varga, {\em Stochastic variational
approach to quantum mechanical few-body problems} (Springer Verlag,
Berlin, Heidelberg, 1998).
\bibitem{isg} N. Isgur and G. Karl, Phys. Rev. D {\bf 18}, 4187
(1978).
N. Isgur and G. Karl, Phys. Rev. D {\bf 20}, 1191 (1979).
\bibitem{sil96} B. Silvestre-Brac, Few-Body Systems {\bf 20}, 1
(1996).
\bibitem{law} R. D. Lawson, {\em Theory of the nuclear shell model}
(Oxford University Press, 1980).
\bibitem{bro} T. A. Brody and M. Moshinsky, {\em Tables of
transformation
brackets} (Monografias del Instituto de Fisica,Mexico, 1960).
\bibitem{sil85} B. Silvestre-Brac, J. Physique {\bf 46}, 1087 (1985).
\bibitem{sema95} C. Semay and B. Silvestre-Brac, Phys. Rev. D {\bf
51}, 1258 (1995).
\bibitem{bhad81} R. K. Bhaduri, L. E. Cohler, and Y. Nogami, Nuovo
Cimento {\bf 65A}, 376 (1981).
\bibitem{fulc94} Lewis P. Fulcher, Phys. Rev. D {\bf 50}, 447 (1994).
\bibitem{isgur} S. Godfrey and N. Isgur, Phys. Rev. D {\bf 32}, 189
(1985); S. Capstick and N. Isgur, Phys. Rev. D {\bf 34},
2809 (1986); L. Ya. Glozman {\em et al.}, Phys. Rev. C {\bf 57},
3406 (1998); L. Ya. Glozman, W. Plessas, K. Varga, and R. F.
Wagenbrunn, Phys. Rev. D {\bf 58},
094030 (1998).
\bibitem{clos79} F. E. Close, {\em An Introduction to Quarks
and Partons} (Academic Press, 1979).
\bibitem{sema97} C. Semay and B. Silvestre-Brac, Nucl. Phys. A {\bf
618}, 455 (1997).
\bibitem{abra70} M. Abramowitz and I. A. Stegun, {\em Handbook of
mathematical functions} (Dover publications, Inc., New York, 1970).
\bibitem{pres92} William H. Press, Saul A. Teukolsky,
William T. Vetterling, and Brain P. Flannerey, {\em Numerical Recipes
in FORTRAN} (Cambridge University Press, 1992).

\end{thebibliography}
\end{document}